\documentclass[preprint,showpacs,prb,preprintnumbers,amsmath,amssymb]{revtex4}
\tighten
\usepackage{graphicx}% Include figure files
\usepackage{dcolumn}% Align table columns on decimal point
\usepackage{bm}% bold math
\begin{document}
\title{The dynamics of superfluid $^4$He}
\author{W. Montfrooij}
\affiliation{Department of Physics and the Missouri Research Reactor, University of Missouri, Columbia Missouri, USA}
\author{E.C. Svensson}
\affiliation {National Research Council Canada, Chalk River Laboratories, Chalk
River Ontario, Canada}
\author{R. Verberg}
\affiliation {Chemical Engineering Department, University of Pittsburgh, Pittsburgh Pennsylvania, USA}
\author{R.M. Crevecoeur and I.M. de Schepper}
\affiliation {Interfaculty Reactor
Institute, Mekelweg 15, 2629 JB Delft, The Netherlands}
\begin{abstract}
We present neutron scattering
  results for the dynamic response by superfluid and normal-fluid $^4$He
  and the results of a simple perturbation-theory analysis which allows us
  to describe all aspects of the observed 
  behavior by considering only density fluctuations.
  We show that the  three key features that are characteristic 
  of the dynamic reponse of superfluid
  $^4$He, as well as their dramatic variations with temperature, 
  can be attributed predominantly  
  to the Bose statistics obeyed by all the 
  $^4$He atoms. It appears that the presence of the Bose condensate exerts at most
  a minor influence on the dynamic response.
\end{abstract}
\date{\today}
\maketitle
\begin{figure}
\centering
\includegraphics[width=15cm]{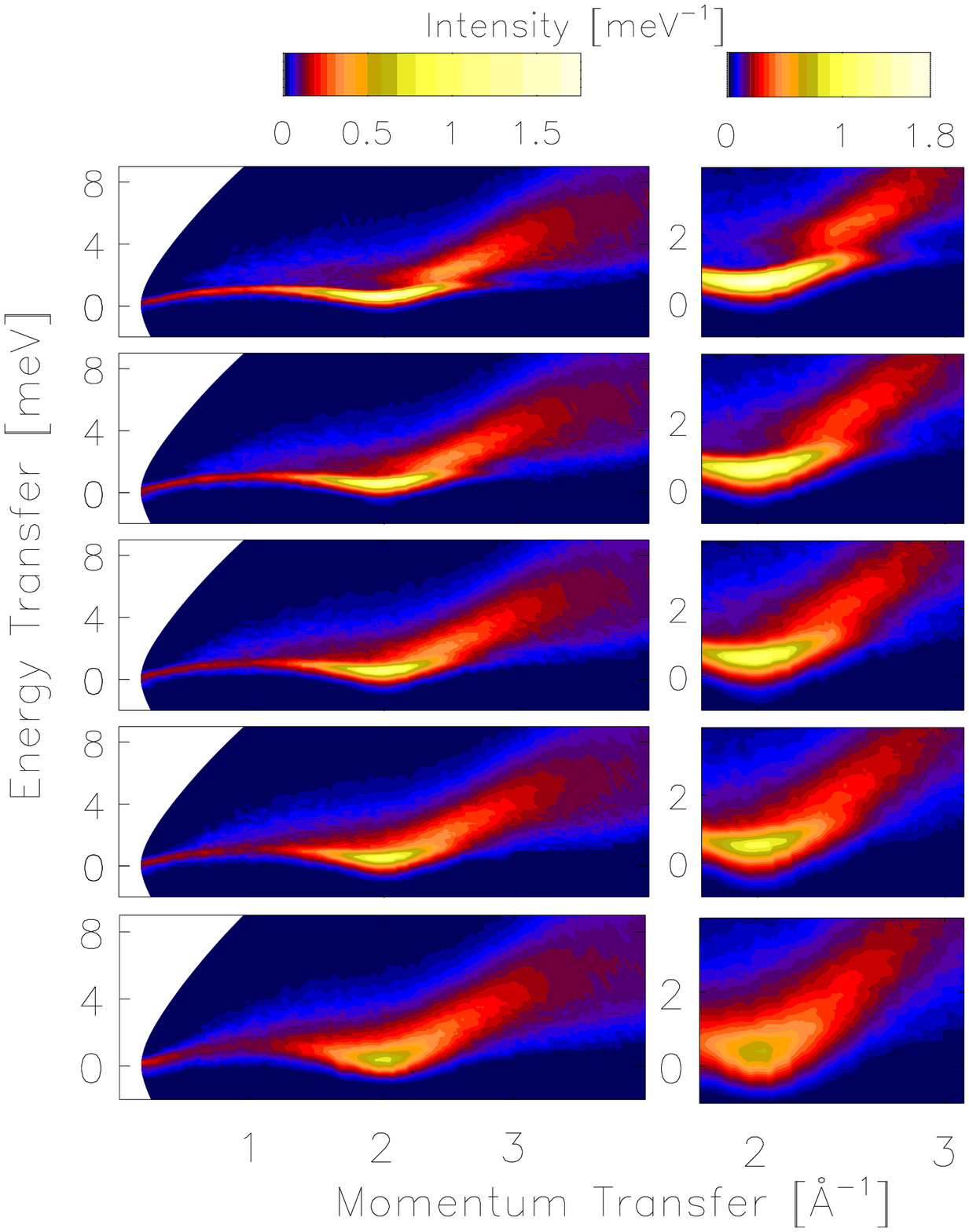}
\caption[]{The (resolution broadened) dynamic response of liquid $^4$He at SVP
for five temperatures ($T$=1.35, 1.85, 2.0, 2.1 and
2.3 K, from top to bottom). We also obtained results (not shown) at 2.15, 
$T_{\lambda}$-0.003 and 3.0 K. The panels on the right show expanded views of the roton
region.} \label{fig2}
\end{figure}
\begin{figure}
\centering
\includegraphics{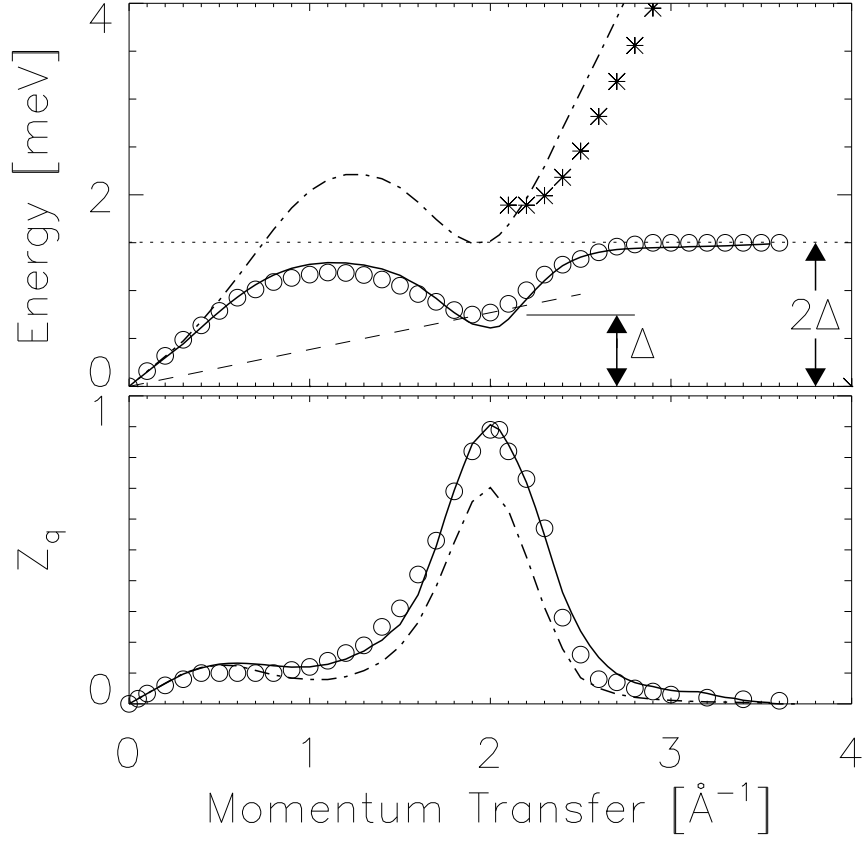}
\caption[]{Top: the phonon-roton dispersion curve of superfluid
$^4$He at $T<$ 1.3 K (circles). The roton excitation 
is given by the minimum of the dispersion curve at $q_r$= 1.92 \AA$^{-1}$,
with energy gap $\Delta$. For $q$-values beyond the roton, 
the dispersion curve asymptotically approaches 2$\Delta$ (dotted line)
and terminates at about 2$q_r$. The stars
indicate the peak positions of the recoil component of the scattering. The
uncorrected excited state 
energies $\hbar\epsilon_F(q)$ are given by the dashed-dotted curve,
and the corrected energies (self-consistent calculation) by the solid line.
Bottom: corresponding weights $Z_q$ of the phonon-roton 
excitations (circles: experiment, solid line: self-consistent calculation,
dashed-dotted line: simple model).} \label{fig1} 
\end{figure}
\begin{figure}
\centering
\includegraphics{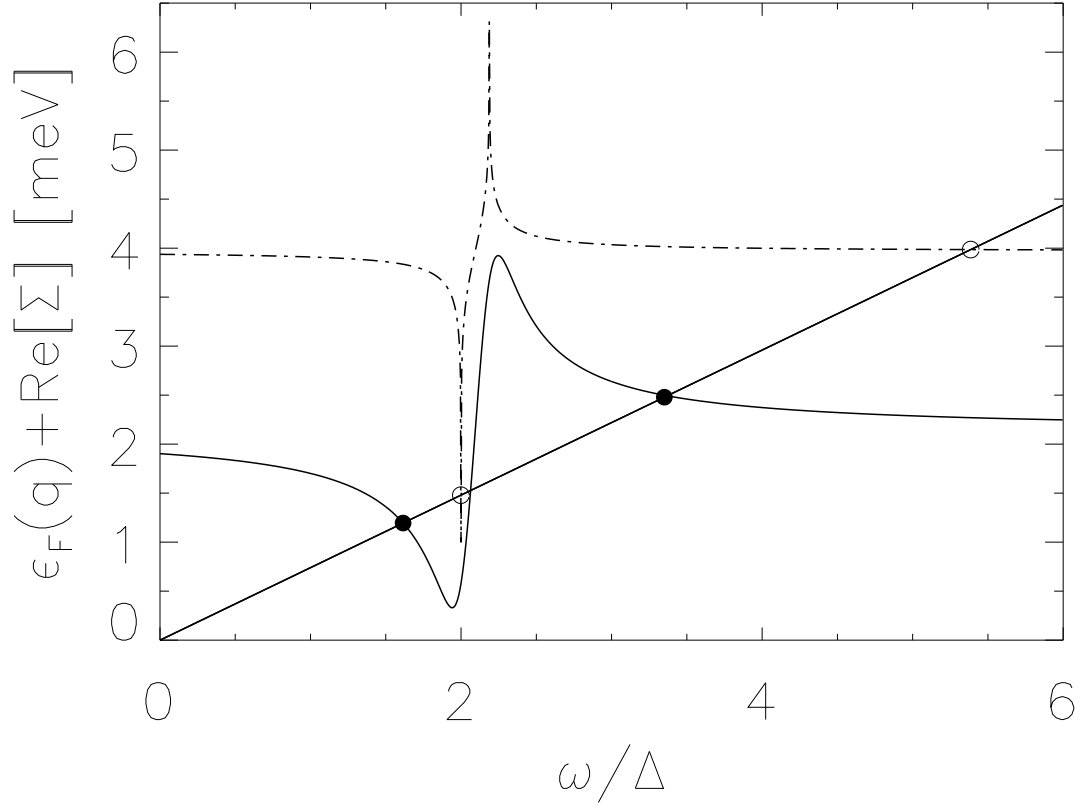}
\caption[]{Solution of the roots of $\omega +i\eta =
\epsilon_F(q)+\Sigma (q,\omega +i\eta )$ for $q$= 2.3 (solid line)
and 2.7$\AA^{-1}$ (dashed line). Only the real part of the
equation is depicted here. In each case there are two solutions, shown by
circles. The apparent third solution does not satisfy the imaginary part of the
equation.} \label{fig3}
\end{figure}
\begin{figure}
\centering
\includegraphics{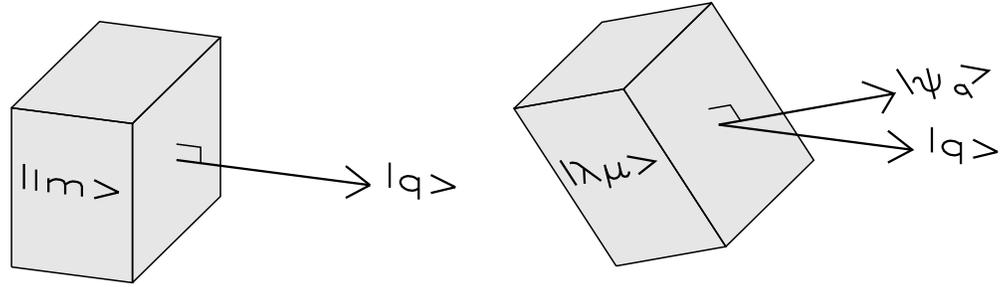}
\caption[]{Pictorial illustration of how an excited (single
quasiparticle) state $|\psi_q>$, which does not coincide with a
simple density fluctuation $|q>$, necessarily leads to the
observation of multi-quasiparticle states $|\lambda \mu>$.}
\label{fig4}
\end{figure}

More than 6 decades after the 
  discovery\cite{kapitza,allen} of
  superfluidity in liquid $^4$He at temperatures 
  below $T_{\lambda}$= 2.17 K, and the almost immediate intriguing proposal\cite{london} that 
  its origin lay in
  the then obscure phenomenon of Bose-Einstein condensation\cite{einstein}, 
  we still do not
  have an all-encompassing theory of superfluid $^4$He. 
The existence of a Bose condensate in superfluid $^4$He has since been confirmed
\cite{sears,wyatt}, and the current\cite{sokol} estimate is
that about 10\% of the
$^4$He atoms are condensed into the zero-momentum state at $T\leq$ 1 K.
  In a recent attempt\cite{gg,griffin,glyde} to develop a microscopic
  theory for liquid $^4$He, the Bose condensate has been deemed to be of central
  importance in determining the dynamic response in the superfluid phase.
  However, key predictions
  of this theory are inconsistent\cite{prl96,icns} 
  with observed behavior.\\ 
On the other hand, the correlated basis function (CBF) approach to the superfluid state, 
which has its
origin in the works of Feynman\cite{feynman} and Jackson and 
Feenberg \cite{jackson_feenberg,jackson} and which 
has since been  
developed by various groups
\cite{campbell,clements1,clements2,apaja,lee,manou}, shows that the 
elementary excitations of superfluid $^4$He can be calculated using a perturbative
expansion based on excited states generated by the density operator. 
Good agreement between
calculated and observed excitation energies has now been established\cite{apaja,lee}.
In this paper, we use the CBF-theory results to forward a unified description
of all three  components in the
dynamic response (elementary excitations, multiphonon and recoil scattering) 
and the variations with temperature thereof.\\
%Liquid $^4$He has been studied more extensively by neutron scattering than any
%other material, with some of the very earliest 
%measurements\cite{palevsky,yarnell} of the dynamics
%Early neutron scattering experiments
%\cite{palevsky,yarnell} 
%on $^4$He 
%confirmed Landau's\cite{landau} famous phonon-roton dispersion relation for
%the elementary excitations, 
%a modern version\cite{donnelly,henry} of
%which is shown in Fig. \ref{fig1}. 
Ever since the Landau's\cite{landau} prediction of the 
famous phonon-roton dispersion relation for
the elementary excitations,  a large body of inelastic neutron scattering work has been accumulated
(see \cite{griffin,glyde} for references), 
revealing the unique features of superfluid $^4$He in ever
increasing detail. Using the
MARI spectrometer at the ISIS pulsed neutron facility, 
we have extended recent measurements\cite{fak,ken,henry} to cover a larger dynamic and
temperature range, focussing on the region where the excitation curve splits
up into three branches. Results for the dynamic response functions
at saturated vapor pressure (SVP)  are shown in Fig. \ref{fig2}.\\
At the lowest
temperature ($T$= 1.35 K, top panels
of Fig. \ref{fig2}), the spectra show three distinct
features characteristic of the superfluid phase.
First, one observes the phonon-roton
dispersion curve with the very intense roton 
excitations seen
as the bright yellow region in Fig. \ref{fig2}.
We show (a compilation\cite{henry,donnelly,new_us} of) the energies and spectral weights of these  excitations in Fig. \ref{fig1}. 
Their lifetimes (related to the
reciprocal of their intrinsic energy widths in the neutron spectra) are
determined by the number of thermally excited
quasiparticles\cite{landau-khalatnikov} present at that temperature,
which in turn govern
the frequency of collisions between the quasiparticles. At $T$= 1.35 K,
the observed widths primarily reflect  the instrumental resolution function (FWHM$<$ 0.25 meV).
Note also that there is no scattering at energies
below the phonon-roton dispersion curve.\\
Second, as a reddish band above and beyond the roton region in Fig. \ref{fig2}
(and stars in Fig. \ref{fig1}),
one observes a broad (in energy) response centred near the
recoil energy $\hbar\epsilon_r=\hbar^2 q^2/2m$, clearly separated form the scattering 
on the 
phonon-roton dispersion curve. This component of the
response, from which the Bose condensate fraction can be extracted\cite{sears,sokol}, 
mainly reflects the behavior of the individual $^4$He
atoms.\\
Third, there is the 
possibility of a neutron  simultaneously
exciting multiple quasiparticles. This multiphonon  can 
be observed as a band of (weak) intensity located
just above the phonon-roton dispersion curve and merging into the recoil
scattering at larger $q$.\\
In contrast, the normal-fluid dynamic
response (bottom panels of Fig. \ref{fig2}) does not
show any sharp excitations, nor is there any separation 
between
the high energy recoil scattering and the low energy response. In addition,
line-shape analysis of these and earlier\cite{icns} results indicates that the
multiphonon component is absent, or at least vanishingly small
in the normal-fluid phase. We first discuss how the features of the
spectra at the lowest temperature can be understood using (only)
perturbation theory and Bose statistics.\\
The shape of the phonon-roton
dispersion curve is sufficient to ensure that the fluid is a superfluid. 
Specifically, the finite energy gap (roton-minumum at $q_r$= 1.92
\AA$^{-1}$; gap $\Delta$= 0.74 meV= 8.6 K) means that
a minimum velocity (given by the slope of the dashed line in the
top panel of Fig. \ref{fig1}) is required in
order to excite a roton out of the ground state, with
frictionless superflow for lower velocities. 
%Note that, 
%due to the creation of vortices,  
%the actual critical flow velocity of $^4$He through a macroscopic capillary is 
%only a fraction of what is expected for the above mentioned minimum velocity 
%of about 60 m/s.
%However, flow velocities for $^4$He through micro-orifices\cite{varoquaux}, 
%combined with the
%results for experiments in which vortices are created using negative 
%ions \cite{hendry} 
%have demonstrated that critical velocities as high as 60 m/s do indeed exist.\\
As Feynman
pointed out \cite{feynman}, the presence of the energy gap is a
direct consequence of the symmetry requirements placed on the
wavefunction by the Bose statistics obeyed by the $^4$He atoms. This
energy gap also ensures that there can be only a few thermally
excited quasiparticles at the lowest temperatures, which in turn
leads to the long lifetime of the quasiparticles because of the
vastly reduced collision frequency between
them\cite{landau-khalatnikov}. This explains the sharpness of the
phonon-roton excitations, and, by consequence, also the full 
set of features observed in 
the dynamic
response function, as we shall now demonstrate.\\
Using perturbation theory, it is straightforward to
derive a shape for the spectral response consistent with the observed spectra. 
Following Feynman\cite{feynman}, we
start with a variational estimate for the excited state $|q)$ by
letting the density operator $\rho_q$, which creates
a density fluctuation of wavelength 2$\pi$/$q$ somewhere in
the liquid, act on the ground
state: $ |q)=\rho_q^+|0> $. The energy of this state is 
$ \hbar\epsilon_F(q)= (q|H|q)/(q|q)= \hbar^2 q^2/2mS_q $, 
with $S_q$ being the static structure factor.\\ 
We correct this state by including the two-quasiparticle states
$|lm)=\rho_l^+\rho_m^+|0>$. In the following, we use the symbols 
$|q>$ and $|lm>$ 
to indicate that the states have been orthonormalized.
The corrected
excited state energies are now given by the
poles of the corresponding Green function $G_1$, defined by
\begin{equation}
G_1=[\omega -\epsilon_F(q)-\Sigma (q,\omega )]^{-1}.
\label{pooltjes}
\end{equation}
Given suitable expressions for the interaction matrix elements $<q|H|lm>$
present in the self-energy $\Sigma $,
the above equation can be solved in an iterative manner, and good
agreement
with the observed phonon-roton dispersion curve has been obtained for $q < 3 \AA^{-1}$
\cite{apaja,lee}.\\
The validity of the above set of equations is readily demonstrated by performing a self-consistency check.
We calculate the $<q|H|lm>$ using a combination\cite{jackson_feenberg,lee} 
of convolution (for 
$q<2 \AA ^{-1}$) and superposition approximations ($q>2 \AA ^{-1}$).
The self-energy is calculated at $T$= 0 K using the observed excitation energies 
$\epsilon_{obs}$ as
input, yielding the corrected excitation energies $\epsilon_F(q)-\Sigma (q,\omega=
\epsilon_{obs}(q))$ [which ideally would coincide with $\epsilon_{obs}(q)$]. We
plot these corrected energies, together with their corresponding intensities, 
as solid lines in Fig. \ref{fig1}. Clearly, the agreement with experiment is good, even
near the end of the dispersion where the corrections to $\epsilon_F(q)$ are very large. 
The discrepancies reflect inaccuracies in the
approximations for the matrix elements, and the neglecting of states with more than two
excitations.\\ 
Since
our primary aim in this paper is to
elucidate the basic mechanisms governing the dynamic response of $^4$He instead of
discussing the details of the interaction,
we limit
our discussion in the following to a simplified model, 
which nevertheless still
contains the essential physics. Since the model 
reproduces the correct general
{\it shape} of $\Sigma$, it will prove to be sufficient for 
interpreting the experimental results.
In this model, 
we only consider two-quasiparticle states with momenta
$l$ and $m$ near the roton minimum and we replace the
matrix elements in this region by an effective $M_{eff}(q)$. As before, 
we replace the energy
dependent corrected energies entering the self-energy $\Sigma $ 
by the observed
excitation energies $\hbar\epsilon_{obs}$ and 
lifetimes $\hbar\Gamma$, yielding 
%.Thus we use the following approximation for $\Sigma $:
\begin{equation}
\Sigma (q,\omega )= \frac{1}{2} \sum _{l,m}
\frac{|M_{eff}(q)|^2\delta_{\overrightarrow{q},\overrightarrow{l}+\overrightarrow{m}}} 
{\omega -\epsilon_{obs}(l)-\epsilon_{obs}(m)-2i\Gamma}.
\label{self}
\end{equation}
Determining the poles of 
$G_1$ can now be done by finding the roots of the much simplified
complex equation $\omega +i\eta = \epsilon_F(q)+\Sigma (q,\omega
+i\eta )$.\\
We show the solution in Fig. \ref{fig3} for
$q$= 2.3 and 2.7$ \AA^{-1}$ ($M_{eff}(q)$ was fixed by imposing self-consistency
on the corrected excitation energies).
One observes that the equation yields
two solutions due to the appearance of a strong resonance in the
selfenergy at 2$\Delta$, i.e., at twice the roton energy. 
This
pronounced resonance is a {\it direct} consequence of the high density of
states near the roton minimum in combination with the long lifetimes
of the elementary excitations. For $q$-values beyond the roton
($q\gtrsim 2.5 \AA^{-1}$). Eq. \ref{pooltjes} yields one solution
with excitation energy near $\hbar\epsilon_F(q)$ ($\simeq
\hbar\epsilon_r$ in this region) and one solution which asymptotically approaches
2$\Delta$, almost independent of the actual values of 
$\epsilon_F(q)$ and $M_{eff}(q)$. This is exactly what is 
observed\cite{henry,new_us} in experiment
(see Fig. \ref{fig1}). Note that the linewidth of
the solution near 2$\Delta$ is roughly given by twice the
roton linewidth, as follows directly from 
Eq. \ref{self} since the imaginary part of the denominator is
given by twice the roton linewidth. This linewidth is still very small
at $T <$ 1 K.\\
For lower $q$-values a
similar process takes place, resulting in the
lowering of the excitation energies with respect to 
$\epsilon_F(q)$ (see Fig. \ref{fig1}) and in the hybridization of 
the excitations making up the recoil component and
the phonon-roton excitations  near $q$= 2.3 $\AA^{-1}$. Of
course, the two-roton resonance disappears once $q$ approaches 2$q_r$
(see Eq. \ref{self}). To
illustrate how realistic this simplified model still is, we show the
predicted intensities $Z_q$ as a dashed curve in the 
lower part of in Fig. \ref{fig1}.
The model thus shows, for all the excitations that constitute
the phonon-roton
dispersion curve, that the excitations 
are density fluctuations in
character. They are {\it not} 
the result of knocking an atom
out of the Bose condensate\cite{gg,griffin,glyde}.\\ 
The corrected
excited state vectors are also determined in the standard manner\cite{manou}. We illustrate
in Fig. \ref{fig4} how this gives rise to the appearance of a
multiphonon component. The corrected single quasiparticle
excited states $|\psi_q>$ are obtained by rotating 
$|q>$ in the Hilbert space spanned by [$|q>$, $\{
|lm>\}$]. 
Whereas the uncorrected states $|lm>$ were orthogonal to
$|q>$, the corrected states
$|\lambda\mu>$ are orthogonal to $|\psi_q>$ (thus,
$<\lambda\mu|q>$ $\neq$ 0). Since the neutron scattering cross-section 
of a state $|..>$ is proportional to
$|<..|q>|^2$,
the
multi-quasiparticle states can so be directly populated by neutrons, leading to the observed
multiphonon component.\\
We now briefly summarize the observed changes in the neutron spectra
on increasing the temperature. 
There is an increasingly rapid decrease in the
lifetime of the excitations \cite{icns} as one approaches $T_{\lambda}$
(see Fig. \ref{fig2}).
This can be attributed solely to an increase in the collision frequency 
resulting
from the greater number of thermally excited
quasiparticles\cite{landau-khalatnikov}, there is no need to invoke
depletion of the Bose condensate. The intensity in the
multiphonon component is transferred to the broadened single
excitation\cite{icns}. The
separation between the phonon-roton dispersion curve and the
recoil component first weakens and then disappears (see Fig. \ref{fig2}).
The excitations near the two-roton threshold rapidly
broaden (with at least twice the roton linewidth \cite{henry}) and
weaken in intensity.\\
We can understand the above changes, at least
qualitatively, from the results of the simple model.
When the lifetime of the roton decreases, the two-roton resonance
will become less pronounced. As discussed above, it follows
directly from our model that the
linewidths of the excitations near the end of the phonon-roton dispersion 
curve ($q$
$\gtrsim$ 2.5 \AA$^{-1}$) increase twice as fast as the roton
linewidth increases. In addition, by repeating the model calculations
using increased roton linewidths, it follows that their intensities
diminish with increasing linewidth so that these
excitations can no longer be separated from the recoil component
\cite{henry} (see Fig. \ref{fig2}). Also, with the weakening of the
two-roton resonance, the excited state energies will be closer to
$\hbar\epsilon_F(q)$. As a
consequence, the excited states $|\psi_q>$ will more closely
resemble a simple density fluctuation $|q>$, meaning that the
multiphonon states $|\lambda\mu>$ will have a smaller
weight (or even no weight at all) in the neutron scattering
spectra (see Fig. \ref{fig4}). The weakening of the two-roton resonance is
probably also the reason why the hybridization weakens (since it
is the origin of the hybridization in the first place), but the
simple model we have used here is not sufficiently accurate to 
allow one to follow
these effects into the region close to the superfluid transition where the
damping rates become comparable to the excitation
energies\cite{icns}.\\
In summary, the results of our inelastic
neutron scattering experiments and our calculations using
perturbation theory are in good qualitative agreement. 
All three key features of the observed 
spectra in the superfluid phase 
are fully explained by the long
lifetimes of the elementary excitations, and the changes 
on increasing the temperature can be attributed to the
weakening of the two-roton resonance, the latter having its origin
in the decreasing lifetime of the roton excitations. Our conclusion is
that the dynamic response of superfluid $^4$He is determined
(primarily) by density fluctuations and that,
although a Bose condensate exists in superfluid $^4$He, it plays 
at most a minor role in determining this response. \\
\acknowledgements 
We thank the technical staff at ISIS for their expert
assistance during the experiments. This research was supported by
the Netherlands Organization for Scientific Research (NWO).\\
 
\clearpage 
\clearpage 
\clearpage 
\clearpage 
\clearpage
 

\begin{thebibliography}{10}
\bibitem{new_us} Our experiments corroborate Ref. \cite{henry} in that the phonon-roton excitation energy never exceeds $2\Delta$ upto $q= 3.6 \AA^{-1}$.
\end{thebibliography}
\end{document}